# Atomic reconstruction and moiré patterns in transition metal dichalcogenide van der Waals heterostructures

Matthew R. Rosenberger,*[#1] Hsun-Jen Chuang,*[%1] Madeleine Phillips,[#1] Vladimir P. Oleshko,[2] Kathleen M. McCreary,[1] Saujan V. Sivaram,[#1] C. Stephen Hellberg,[1] and Berend T. Jonker[1]

* These authors contributed equally to this work.

[1] U.S. Naval Research Laboratory, Washington, DC

[2] National Institute of Standards and Technology, Gaithersburg, MD

[#] National Research Council Postdoctoral Fellow

[%] American Society of Engineering Education Postdoctoral Fellow

Van der Waals layered materials, such as transition metal dichalcogenides (TMDs), are an exciting class of materials with weak interlayer bonding which enables one to create so-called van der Waals heterostructures (vdWH).[1] One promising attribute of vdWH is the ability to rotate the layers at arbitrary azimuthal angles relative to one another. Recent work has shown that control of the twist angle between layers can have a dramatic effect on vdWH properties, including the appearance of superconductivity,[2,3] emergent electronic states,[4–7] and unique optoelectronic behavior.[6–11] For TMD vdWH, twist angle has been treated solely through the use of rigid-lattice moiré patterns. No atomic reconstruction, i.e. any rearrangement of atoms within the individual layers, has been reported experimentally to date. However, any atomic level reconstruction can be expected to have a significant impact on the band structure and all measured properties, and its existence will fundamentally change our understanding and theoretical treatment of such systems. Here we demonstrate that vdWH of MoSe$_2$/WSe$_2$ and MoS$_2$/WS$_2$ at twist angles ≤ 1° undergo significant atomic level reconstruction leading to discrete commensurate domains divided by narrow domain walls, rather than a smoothly varying rigid-lattice moiré pattern as has been assumed in prior work. Using conductive atomic force microscopy (CAFM), we show that the stacking orientation of the two TMD crystals has a profound impact on the atomic reconstruction, consistent with recent theoretical work on graphene/graphene and MoS$_2$/MoS$_2$ structures at small angular misalignments[12] and experimental work on graphene bilayers[13–15] and hBN/graphene heterostructures.[16,17] Transmission electron microscopy (TEM) provides additional evidence of atomic reconstruction in MoSe$_2$/WSe$_2$ structures and demonstrates the transition between a rigid-lattice moiré pattern for large angles and atomic reconstruction for small angles. We use density functional theory (DFT) to calculate the band structures of the commensurate reconstructed domains and find that the modulation of the relative electronic band edges is consistent with the CAFM results and photoluminescence spectra from reconstructed vdWH. The presence of atomic reconstruction in TMD heterostructures and the observed impact on nanometer-scale electronic properties provides fundamental insight into the behavior of this important class of heterostructures.

The individual layers of vdWH possess exceptionally strong *intra*-layer bonding and weak *inter*-layer bonding, with no chemical bonds between the layer planes. The weak interlayer bonding frees one from the constraint of lattice matching, thereby enabling a new approach for "materials by design" by sequentially stacking non-lattice matched monolayers to form unique van der Waals heterostructures, an avenue not possible with traditional epitaxial approaches dominated by out-of-plane bonding.[1] This represents a "bottom up" approach towards design and fabrication of new materials that do not exist in nature, and a new class of atomic-scale heterostructures that are expected to exhibit properties and functionality beyond the limits of their bulk counterparts.

In addition to selection of the stacking sequence, the azimuthal orientation or "twist angle" between such layers can be independently controlled, and provides an additional degree of freedom not present in conventional materials. The superposition of two similar patterns, such as atomic lattices, often gives rise to an additional larger periodicity known as a moiré pattern. A moiré pattern can arise either from a difference in lattice constant of the two patterns or a twist angle between the two patterns. In vdWH formed from materials with markedly different lattice constants, a moiré pattern forms regardless of angular misalignment, as has been demonstrated for TMD vdWH.[18] In vdWH formed from materials with nearly identical lattice constants, such as the vdWH investigated in this work, a moiré pattern will form for any misalignment angle between the two layers.

We fabricated $MoSe_2$/$WSe_2$ and $MoS_2$/$WS_2$ vdWH as described in the Methods. Triangular single-crystal grains of monolayer TMD were grown by chemical vapor deposition (CVD) and the edges of the triangular grains were used for angular alignment of layers during fabrication. We consider two twist angle configurations of interest: 0° alignment in which the triangle vertices are aligned to one another at an angle of 0°+δ, where δ is the misalignment angle as shown in Fig. 1(a) and 60° alignment in which the triangle vertices are aligned to one another at an angle of 60°+δ as shown in Fig. 1(d).

The expected moiré pattern for a 0°+δ and a 60°+δ structure is shown in Fig. 1(a) and Fig. 1(d), respectively. The plots are shown for δ = 4° for clarity, but the pattern is similar for different angles (see Fig. S1). Although the local stacking arrangement (i.e. the relative spacing orientation of atoms in the two layers) varies continuously across the structure, each point in the moiré pattern approximates a stacking arrangement that is obtained by simple translation of one atomic lattice relative to another without any twist angle.[12] Certain points in the moiré pattern correspond to high-symmetry stacking arrangements. In Fig. 1(a), two high-symmetry stacking arrangements are labeled with green and gold circles for the 0°+δ vdWH: AB (transition metal [A] in top layer is above the chalcogen [B] in the bottom layer) and BA (chalcogen in top layer is above the transition metal in the bottom layer). In Fig. 1(d), one high-symmetry stacking arrangement is labeled with red circles for the 60°+δ vdWH: ABBA (transition metal in top layer is above the chalcogen in the bottom layer and the chalcogen in the top layer is above the transition metal in the bottom layer). (See Table S1 for a summary of the different stacking arrangement nomenclature used in the literature.) Importantly, moiré patterns in vdWH, like those shown in Fig. 1(a) and Fig. 1(d), are simply a consequence of the geometry of the rigid lattices and do not account for interlayer interactions.

The significance of the high symmetry points labeled in Fig. 1(a) and Fig. 1(d) are that these stacking arrangements correspond to the minimum interlayer stacking energy in the structures. Figure 1(b) and Fig. 1(e) show plots of the stacking energy as a function of position along the red lines labeled in Fig. 1(a) and Fig. 1(d), respectively. (Figure S2 shows similar calculations for $MoS_2$/$WS_2$ vdWH which exhibit similar behavior to $MoSe_2$/$WSe_2$ vdWH.) From Fig. 1(b), the 0° structure exhibits two degenerate low-energy stacking arrangements: AB and BA. This spatial variation in stacking energy causes a driving force for atomic reconstruction such that the AB and BA domains will grow, as depicted in the inset of Fig. 1(b). Since each AB domain has three nearest neighbor BA points (and vice versa), we anticipate that the AB and BA domains will be triangular, similar to theoretical predictions for 0° $MoS_2$/$MoS_2$ vdWH.[12] In contrast with the 0° case, the 60° structure exhibits a single low-energy stacking arrangement: ABBA [Fig. 1(e)]. Similar to the 0° case, there is driving force for atomic reconstruction, but there will only be one domain type, as depicted in the inset of Fig. 1(e). Since each ABBA point has six nearest neighbor ABBA points, we anticipate that the ABBA domains will be hexagonal, similar to theoretical predictions for 60° $MoS_2$/$MoS_2$ vdWH.[12] The degree to which atomic reconstruction occurs depends on the relative strength of the potential energy gradient due to interlayer stacking energy, which drives atomic reconstruction, and the restoring force due to intralayer strain, which opposes atomic reconstruction.[12]

Figure 1(c) and Fig. 1(f) provide experimental proof that significant atomic reconstruction occurs in $MoSe_2$/$WSe_2$ heterostructures at small twist angles ($\delta \leq 1°$), which is a significant departure from the pattern anticipated from a simple rigid-lattice moiré pattern (i.e. no consideration of interlayer interactions). Figure 1(c) shows a CAFM measurement of a $MoSe_2$/$WSe_2$ vdWH with 0°+δ alignment where we estimate δ = 0.4° from the period of the pattern (see Fig. S3 for details of angle estimation). The measurement reveals a lattice of discrete triangular domains with alternating high and low conductivity, which is not expected from conventional moiré theory. The existence of triangular domains is instead consistent with atomic reconstruction, which produces large commensurate domains of the lowest stacking energy structures, as shown in the inset to Fig. 1(b). The presence of two types of domains with different conductivity agrees with the prediction of two low energy stacking configurations, AB and BA, which our DFT calculations further predict to have different electronic properties (discussed below). Importantly, the AB and BA domains are non-equivalent in $MoSe_2$/$WSe_2$ vdWH because the layers are not the same,[19] which is different from the case of graphene/graphene or $MoS_2$/$MoS_2$ structures.[12] We also observed a similar pattern of alternating conductivity triangular domains in 0°+δ $MoS_2$-$WS_2$ vdWH (Fig. S4).

A strikingly different pattern is observed for the 60° alignment, as shown by the CAFM image in Fig. 1(f) for a similar $MoSe_2$-$WSe_2$ vdWH, but with 60°+δ alignment with estimated δ = 0.5° – 0.8°. The measurement shows large hexagonal domains with constant conductivity separated by domain boundaries of different conductivity, a significant departure from conventional moiré theory in which one expects a continuously varying band structure.[4] The existence of hexagonal domains is instead consistent with atomic reconstruction to form large commensurate hexagonal domains, as shown in the inset of Fig. 1(e). The constant conductivity observed for all domains is consistent with the prediction of a single low-energy stacking configuration for 60° aligned structures.

TEM data corroborate the presence of atomic reconstruction in $MoSe_2$/$WSe_2$ vdWH and demonstrate the transition between simple rigid-lattice moiré patterns at large twist angles and atomic reconstruction at small twist angles. Figure 2(a) shows a schematic of Sample 1, which is a 0°+δ structure with δ = 3°. Figure 2(b) shows a phase contrast high-resolution TEM (HRTEM) image of the structure which reveals a rigid-moiré pattern with no atomic reconstruction domains, similar to the one depicted in Fig. 1(a). Figure 2(c) shows a dark-field (DF-) TEM image of the same structure which reveals one-dimensional fringes that are characteristic of rigid-lattice moiré structures.[20] The diffraction spot circled in the selected-area electron diffraction (SAED) pattern [inset of Fig. 2(c)] was used for the DF-TEM image. Figure 2(d) shows a schematic of Sample 2 with δ = 1°, which consists of a single crystal triangle of $MoSe_2$ transferred on top of a $WSe_2$ flake which has two domains with crystal orientation rotated 60° from one another. The two domains in the $WSe_2$ flake are separated by mirror twin boundaries.[21] Therefore, this sample has both 0° and 60° vdWH immediately adjacent to one another, separated by the $WSe_2$ mirror twin boundaries. Figure 2(e) shows a bright-field (BF-) TEM image spanning both the 0° and 60° regions of the sample in which hexagonal domains are visible on the right side of the mirror twin boundary. Figure 2(f) shows a DF-TEM image of the region labeled with a red box in Fig. 2(e). On the left side of the mirror twin boundary (0° region), there is a lattice of triangular domains with alternating image contrast. The alternating contrast indicates that the two domains have different crystal orientations and/or structures,[13,14] which agrees with the observations from CAFM [Fig. 1(c)]. On the right side of the mirror twin boundary (60° region), there are hexagonal domains separated by domain boundaries. All hexagonal domains have the same signal intensity, indicating that all hexagons have the same crystal structure and orientation, which also agrees with the observations from CAFM [Fig. 1(f)]. HRTEM of a reconstructed sample provides further proof that the reconstructed domains have commensurate stacking [Fig. S6]. We also observe that the average area of hexagonal domains in the 60° region is approximately twice that of the triangular domains in the 0° region, as anticipated from the diagrams in Fig. 1(b) and Fig. 1(e).

As the angular misalignment decreases, the size of reconstructed domains increases. Figure 2(g) shows a schematic of Sample 3, which is similar to Sample 2, except the orientation of the MoSe$_2$ layer is rotated by 60°. Figure 2(h) shows a BF-TEM image from the 0° region, which exhibits a lattice of triangular domains with approximate domain size of 52 nm, from which we estimate δ = 0.4°. In DF-TEM, the triangles again show alternating contrast [Fig. 2(i)]. The shape of the triangular domains is less regular in Sample 3 compared to Sample 2. As a general trend, we observed that the atomic reconstruction pattern is less regular for larger domains. Figure 2(k) shows a BF-TEM image of the 60° region of Sample 3 labeled with a black box in Fig. 2(j). The image shows several hexagonal domains of various sizes ranging from 20 nm up to 120 nm in size. Also, there are domain walls of about 2 nm to 5 nm in width emanating from a small patch of bilayer that was on the MoSe$_2$ layer. We generally found that regions of bilayer, edges of flakes, and tears/rips in the sample led to complex patterns of atomic reconstruction. Similar to the 60° aligned region in Fig. 2(f), all of the hexagonal domains exhibit the same contrast in DF-TEM [Fig. 2(l)].

The CAFM data, TEM data, low-energy stacking sequences identified by DFT, and the atomic reconstruction theory[12] all converge on the following picture: atomic reconstruction results in commensurate triangular domains with AB and BA stacking arrangements for 0° structures with δ ≤ 1° and atomic reconstruction results in commensurate hexagonal domains with ABBA stacking arrangement for 60° structures with δ ≤ 1°.

The presence of atomic reconstruction has significant implications for the electronic and optical properties of TMD vdWH. Figure 3(a-b) show conductive AFM measurements of both 0° and 60° heterostructures with (a) sample bias, $V_s > 0$ and (b) $V_s < 0$ (b). The images have been flattened for clarity (raw data in Fig. S7). The sample geometry was similar to the samples used in Fig. 2, yielding both heterostructure orientations in the same sample. The observed patterns are qualitatively the same for both $V_s > 0$ and $V_s < 0$. Namely, there are hexagonal domains of varying sizes on the left half of the images (60° region) and various sizes of triangles of alternating conductivity on the right half of the images (0° region). The same triangles maintain higher conductivity for both $V_s > 0$ and $V_s < 0$. We hypothesize that variation in size and aspect ratio of the hexagons and triangles indicates local variations in strain.[22] Figure 3(c) shows a closeup of a single hexagon from the 60° region. The domain walls in the 60° region exhibit two parallel bands, one band with higher conductivity than the hexagonal domains (labeled with red boxes) and one band of lower conductivity than the hexagonal domains (labeled with gray boxes). Also, the vertices of the hexagon alternate between high conductivity (red circles) and low conductivity (gray circles). This complex domain wall behavior is consistent with the complex atomic arrangements at domain walls predicted by theory[12] and seen experimentally in graphene-graphene structures.[13]

To understand the nature of CAFM contrast and to assign structures to the different domains, we used DFT to calculate the band structures of the low-energy stacking configurations. A simplified picture of the conduction band minimum (CBM) and the valence band maximum (VBM) for each low-energy stacking configuration is shown in Fig. 3(d). Full band structures are shown in Fig. S8. From the DFT calculations, the CBM for BA stacking is lower than the CBM for AB stacking and the VBM for BA stacking is higher than the VBM of AB stacking. This is consistent with BA stacking being more conductive for both $V_s > 0$ and $V_s < 0$ because sample bias will translate the position of the vdWH band structure relative to the fermi level of the CAFM tip. $V_s > 0$ corresponds to the fermi level of the CAFM tip moving toward to the conduction band, such that $V_s > 0$ primarily samples the conduction band. $V_s < 0$ corresponds to the fermi level of the CAFM tip moving toward the valence band, such that $V_s < 0$ primarily samples the valence band. This interpretation is similar to the conventional interpretation of scanning tunneling microscopy dI/dV curves. It follows that the higher conductivity triangles have BA structure and the lower conductivity triangles have AB structure. One high conductivity triangle (BA) is labeled in gold and one low conductivity triangle (AB) is labeled in green in Fig. 3(a-b). We assign ABBA structure to all hexagons in the 60° region because this is the single low-energy stacking configuration. One hexagon is labeled in red in Fig. 3(a-b). General agreement between CAFM experiments and DFT band structure calculations suggests that DFT band structure calculations capture the qualitative physics of these heterostructures.

With an understanding of the electronic structure of the $MoSe_2$-$WSe_2$ vdWH from DFT, we turn to investigating the interlayer optical behavior of the structures. Figure 3(e) shows a spatial map of the wavelength of maximum PL intensity in the range from 900 to 1000 nm (the emission range of interlayer excitons)[23–26] at room temperature for Sample 3 from Fig. 2. Figure 3(f) shows a typical spectrum for the 0° structure and the 60° structure at room temperature. From Fig. 3(e) and Fig. 3(f), there is a clear change in peak emission wavelength for the 0° structure as compared to the 60° structure. The 0° structure exhibits an average wavelength of 961 nm and the 60° structure exhibits an average wavelength of 926 nm. The lower energy (longer wavelength) of the interlayer exciton in the 0° structure as compared to the 60° structure is consistent with the calculated DFT bandgaps, shown in Fig. 3(d). Namely, the smallest bandgap in the 0° structure (BA) is smaller than the bandgap of the 60° structure (ABBA), and the difference in the bandgap of these two structures from DFT (22 meV) compares well with the observed difference in PL emission energy (29 meV). This suggests that the interlayer exciton behavior of vdWH with atomic reconstruction can be understood by considering the band structure of the commensurate domains without considering an additional moiré potential.

In summary, the rigid-lattice moiré construction typically used as a starting point to calculate or interpret the properties of a TMD vdWH is not appropriate for the case of small twist angles (≤ 1°). Instead, our findings indicate that atomic reconstruction is significant for nearly lattice-matched TMD/TMD vdWH at small twist angles. We believe that such reconstructions will occur in any bilayer system with clean interfaces in which the energy gained from adopting low energy vertical stacking configurations is larger than the accompanying strain energy. Such atomic level reconstruction can be expected to have a significant impact on the band structure and all measured properties, and its existence will fundamentally change our understanding and theoretical treatment of such systems. Future efforts to understand and manipulate similar heterostructures should account for atomic reconstruction in order to precisely describe these systems.

**Methods**

*Chemical Vapor Deposition Growth*

Monolayer $WSe_2$ and $MoSe_2$ were synthesized *via* atmospheric pressure CVD using solid precursors in a 5.1 cm quartz tube furnace. Silicon wafers with 275 nm thermally grown oxide (Silicon Valley Microelectronics, Inc.)[&] were used as the growth substrates. Prior to growth, the substrates are cleaned by ultrasonication, piranha etching, and oxygen plasma exposure. $WSe_2$ synthesis used a water-soluble seeding promoter, perylene-3,4,9,10-tetracarboxylic acid tetrapotassium salt (PTAS), that was drop-cast onto a clean $SiO_2$ substrate immediately before growth. A clean $SiO_2$ substrate was placed downstream from the PTAS substrate. The substrates were loaded face down on a quartz boat directly above the solid precursor (1000 mg of $WO_3$ (Alfa Aesar, 99.998%)[&] or 50 mg $MoO_2$(Sigma Aldrich, 99%)[&]). The precursor and substrates were then moved to the center of the tube furnace. An additional quartz boat containing Se (Alfa Aesar, 99.999%)[&] was placed upstream near the edge of the furnace. The tube was repeatedly evacuated to ~13.3 Pa and filled with UHP Ar and $H_2$. The furnace temperature was held at 825 °C for 10 minutes and then cooled to room temperature with Ar (100 sccm) and $H_2$ (10 sccm) flowing continuously. For the $MoSe_2$ synthesis, the freshly cleaned $SiO_2$/Si was placed face down over the $MoO_2$. This boat was placed in the center of the furnace and 500 mg Se was placed upstream in a separate boat at the edge of the furnace. After repeated evacuations with Ar, the furnace was heated to 800 °C at a rate of 20 °C/min with an Ar flowrate of 20 sccm. Upon reaching 800 °C, $H_2$ was added at 2.5 sccm and the temperature was held constant for 10 minutes. The furnace was slowly cooled to 750 °C at a rate of 1.7 °C/min and then rapidly cooled to room temperature.

*Sample Preparation*

All vdWH stacks (MoSe$_2$-WSe$_2$-hBN for TEM and MoSe$_2$-WSe$_2$-graphite for CAFM) samples were fabricated by water-assisted-pick-up transfer method[27] along with the nano-squeegee technique.[28]

**For the CAFM samples:**

First, a monolayer CVD MoSe$_2$ single crystal was picked up by PDMS stamp, then aligned and transferred onto a monolayer CVD WSe$_2$ single crystal followed by the nano-squeegee[28] to create a clean interface between the TMD layers. Second, the MoSe$_2$-WSe$_2$ stack was picked up and transferred onto the desired graphite flakes (mechanically exfoliated) followed by nano-squeegee.[28] Finally, the MoSe$_2$-WSe$_2$-graphite stack was picked up and transferred onto Au (e-beam evaporated 200nm) coated SiO$_2$/Si substrate.

**For the TEM samples:**

The hBN flake (mechanically exfoliated; thickness around 15nm) was used instead of bottom graphite mentioned in the procedure above. MoSe$_2$-WSe$_2$-hBN stack was then picked up and transferred onto TEM grids. For sample 1, the TEM grid was a Cu mesh microscopy grid covered with holey thin amorphous carbon flat film (1.2µm pore diameter). For Sample 2, the TEM grid was holey amorphous SiN (200nm thick, 5µm pore diameter) coated with 50 nm of Au. For Sample 3, the TEM grid was holey amorphous SiN (200 nm thick, 0.5µm pore). The details of the TEM grids are given in the following TEM section.

*Atomic Force Microscopy*

We performed CAFM measurements on a Keysight 9500 AFM[&] using conductive diamond coated cantilevers (AIO-DD from Budget Sensors)[&]. The AFM sample chamber was continually purged with nitrogen and the relative humidity was approximately 3%. Bias applied to the sample backplane (gold) caused current flow between the backplane and the tip. A current amplifier connected to the tip recorded the current. In order to improve spatial resolution, the applied tip load was reduced as much as possible in order to reduce the tip-sample contact area.

*Photoluminescence*

We performed PL measurements at room temperature in a scanning confocal microscope (Horiba LabRam Evolution)[&] using a 532 nm continuous-wave laser.

*Transmission electron microscopy (TEM)*

BF-TEM, DF-TEM, SAED, and phase contrast HRTEM of the specimens were performed in a Titan 80-300 (Thermo Fischer Scientific (FEI), Hillsboro, OR)[&] equipped with a Schottky-type electron gun and S-TWIN objective lens. The instrument was operated at 300 kV accelerating voltage with a point-to-point resolution of 0.19 nm and information limit below 0.1 nm. The images were acquired using a Gatan[&] Ultrascan CCD camera and Digital Micrograph software with exposure times of 1.0-4.0 s with frame sizes of 2048 × 2048 pixels, and electron doses below $3 \cdot 10^6$ e-/nm$^2$.

Information for TEM grids:

- Sample 1 - Ted Pella,[&] Quantifoil® Orthogonal Array of 1.2µm diameter holes 1.3 µm Separation on 200 mesh Cu grid
- Sample 2 - Ted Pella,[&] ELCO® Holey Silicon Nitride Support Film, 200nm, 5µm pores, 0.5 x 0.5mm window, Ø3mm

- Sample 3 - Ted Pella,[&] ELCO® Holey Silicon Nitride Support Film, 200nm, 0.5µm pores, 0.5 x 0.5mm window, Ø3mm

*Density Functional Theory Calculations*

We use the generalized gradient approximation (GGA)[29] and the projector augmented wave (PAW) potentials[30,31] implemented in VASP.[32] For the selenide bilayers, we use metal potentials with 14 valence electrons and selenide potentials with 6 valence electrons. For the sulfide bilayers, the metal and sulfur potentials we use each has 6 valence electrons. For both materials, each calculation is done with an 8x8x1 Γ-centered k-point mesh and a plane wave energy cutoff of 450 eV. For the $MoSe_2/WSe_2$ calculations, an in-plane lattice constant a=3.293 Å was used for the 60° structures, and a=3.295 Å was used for the 0° structures. For the $MoS_2/WS_2$ calculations, an in-plane lattice constant a=3.293 Å was used for the 60° structures, and a=3.295 Å was used for the 0° structures. The out-of-plane lattice constant is 30 Å for all systems, resulting in about 20 Å of vacuum, with the exact amount depending on interlayer spacing. The van der Waals interaction between layers is modeled using the DFT-D3 method of Grimme.[33] To obtain the stacking energies, we relaxed each structure with the chalcogens allowed to relax in all three cartesian directions and the metals fixed in the x-y plane and allowed to relax along the layer normal (z).

[&]Certain equipment, instruments or materials are identified in this paper in order to adequately specify the experimental details. Such identification does not imply

recommendation by the National Institute of Standards and Technology nor does it imply the materials are necessarily the best available for the purpose.

**Acknowledgements**

The authors gratefully acknowledge funding support for this work from core programs at NRL and the NRL Nanoscience Institute. This research was performed while M.R.R., M.P., and S.V.S. held a National Research Council fellowship and H.-J.C. held an American Society for Engineering Education fellowship at NRL.

# Figures

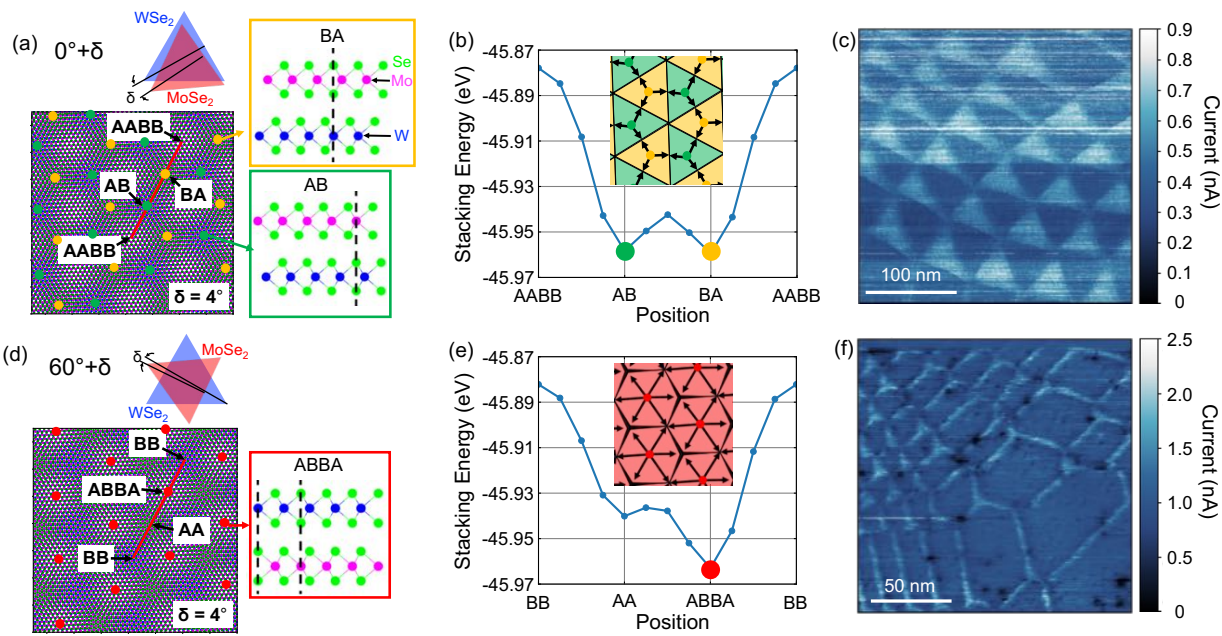

Figure 1. Observation of atomic reconstruction in MoSe$_2$/WSe$_2$ heterostructures. (a) Diagram defining the 0°+δ heterostructure. The image below the diagram shows the moiré pattern formed between the MoSe$_2$ and WSe$_2$ lattices. The moiré pattern is shown for δ = 4° to aid visualization, but the moiré pattern is similar for smaller δ (see Fig. S1). The locations in the moiré pattern which exhibit AB structure (transition metal [A] in top layer above chalcogen [B] in bottom layer) and BA structure (chalcogen in top layer above transition metal in bottom layer) are labeled with green and gold circles, respectively. (b) Interlayer stacking energy as a function of stacking orientation along the red line labeled in (a). AB and BA structure are degenerate global minima, which suggests that AB and BA regions will grow to form larger domains. A diagram indicating the direction of the expected AB and BA domain expansion is shown in the inset. (c) Conductive AFM (CAFM) image of a 0°+δ heterostructure showing alternating triangular domains of different conductivity. We estimate δ = 0.4° for this structure (see Fig. S3). The existence of two distinct types of domains is consistent with the prediction of atomic reconstruction of the lattice two degenerate low energy stackings shown in (b). (d) Diagram defining the 60°+δ heterostructure with δ = 4°. The image below the diagram shows the moiré pattern formed between the MoSe$_2$ and WSe$_2$ lattices. The locations in the moiré pattern which exhibit ABBA structure (transition metal in top layer above chalcogen in bottom layer and chalcogen in top layer above transition metal in bottom layer) are labeled with red circles. (e) Interlayer stacking energy as a function of stacking orientation along the red line labeled in (c). ABBA structure is the global minimum, which suggests that ABBA regions will grow to form larger domains. A diagram indicating the direction of the expected ABBA domain expansion is shown in the inset. (f) CAFM image of a 60°+δ heterostructure showing hexagonal domains of the same conductivity separated by domain boundaries of different conductivity. We estimate δ = 0.5 - 0.8° for this sample. The image has been flattened for clarity (see Fig. S5 for raw data).

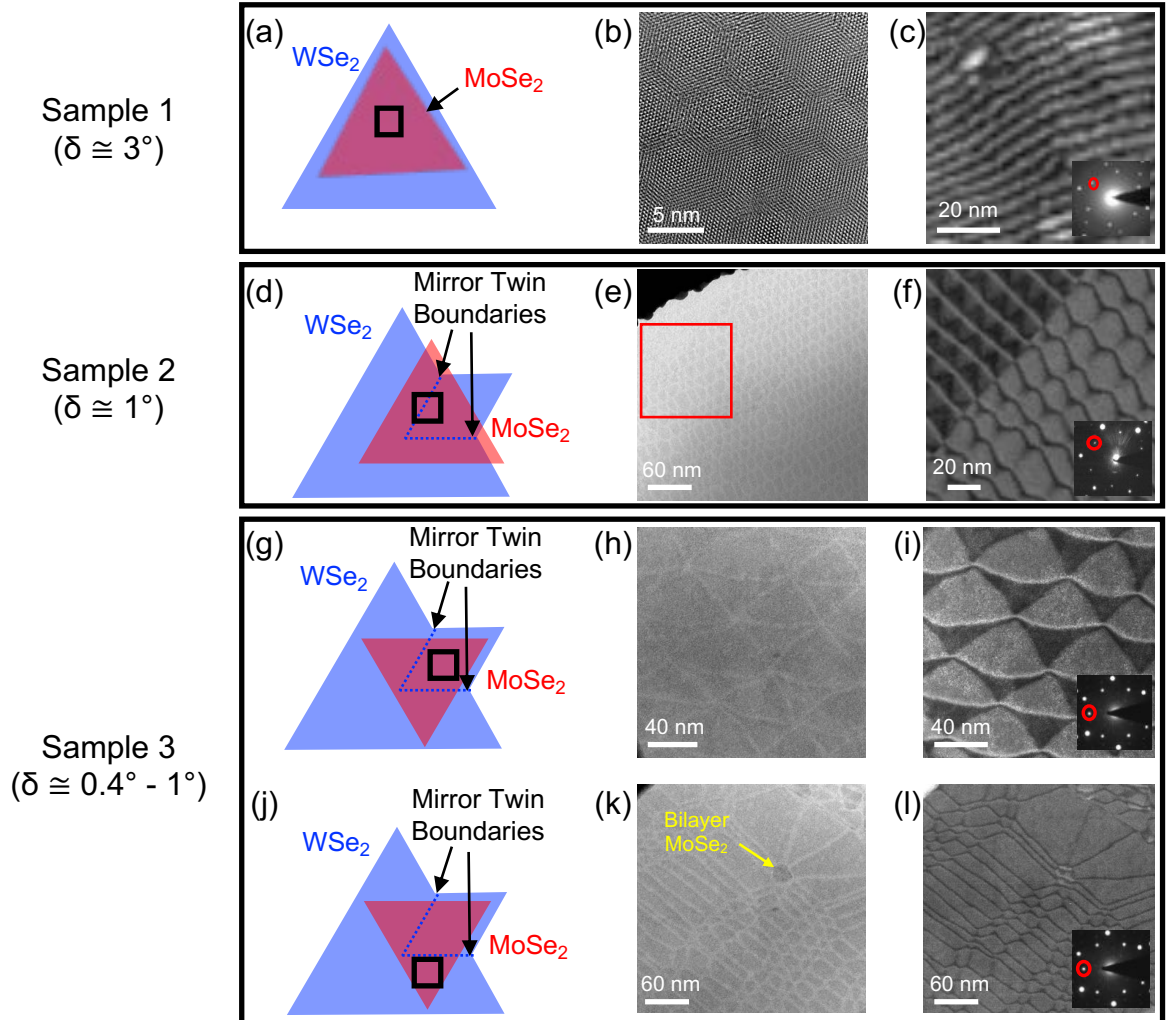

Figure 2. Transmission electron microscopy (TEM) of rigid-lattice moiré and atomic reconstruction in MoSe$_2$/WSe$_2$ heterostructures. (a) Schematic for Sample 1. The black boxes in all sample schematics show approximate location of TEM images. (b) Phase contrast HRTEM image of Sample 1 showing no commensurate domains and a smoothly varying pattern as expected for a rigid-lattice moiré pattern [Fig. 1(a)]. (c) DF-TEM image using the diffraction spot labeled in the inset showing approximately one-dimensional fringes, which are indicative of a rigid-lattice moiré pattern. (d) Schematic for Sample 2. A monolayer MoSe$_2$ triangle was transferred on top of a WSe$_2$ monolayer with mirror twin boundaries, resulting in both 0°+δ and 60°+δ heterostructures in the same sample with the same misalignment angle, δ. (e) BF-TEM of the mirror twin boundary separating the 0°+δ and 60°+δ regions. (f) DF-TEM of the region labeled with a red box in (e) using the diffraction spot labeled in the inset. The image shows a sharp transition between the 0°+δ region (triangles) and the 60°+δ region (hexagons). (g) Schematic for Sample 3 with a 0°+δ region labeled with a black box. (h) BF-TEM image of the 0°+δ heterostructure showing triangular domains. (i) DF-TEM image of the 0°+δ heterostructure showing alternating triangular domains of different atomic arrangement. (j) Schematic of Sample 3 with the black box indicating location of a 60°+δ region. (k) BF-TEM image of the 60°+δ region showing hexagon-like domains. (l) DF-TEM showing that all hexagon-like domains in the 60°+δ heterostructure have the same atomic arrangement.
21

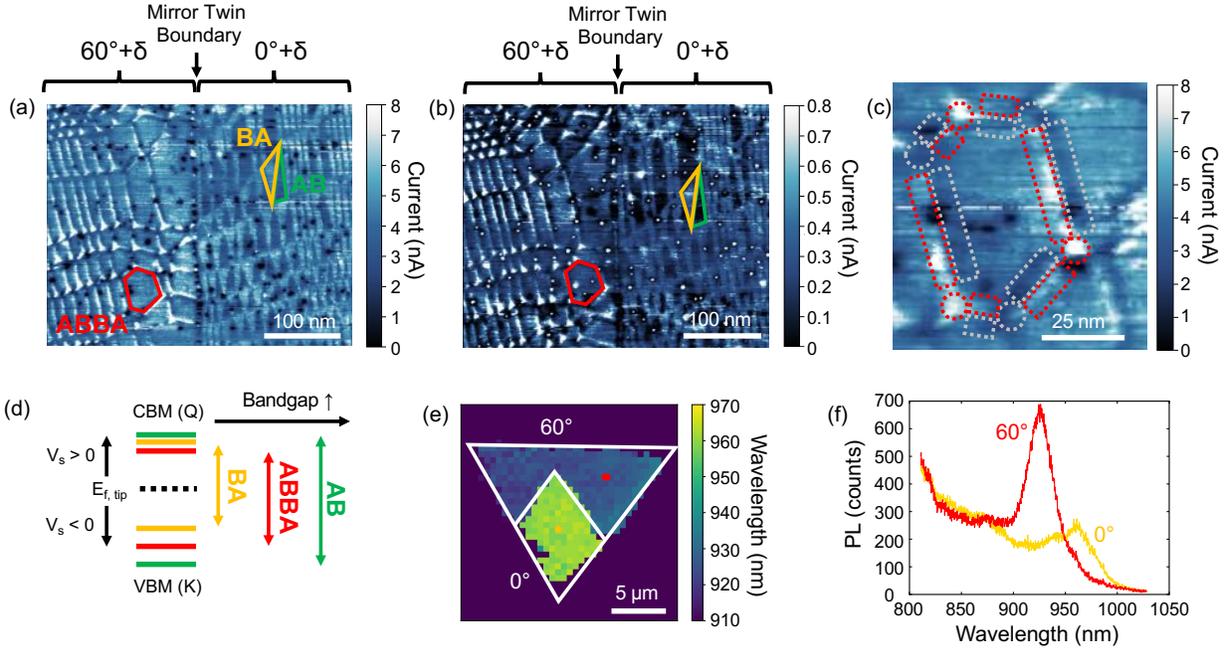

Figure 3. Correlation of bias dependent conductive AFM (CAFM), density functional theory (DFT) band structure calculations, and optical spectroscopy. (a-b) CAFM image of a MoSe$_2$/WSe$_2$ heterostructure on graphite with (a) sample bias, V$_s$ = +0.8 V and (b) V$_s$ = -1 V. The heterostructure is comprised of a single grain of MoSe$_2$ and a WSe$_2$ sample with a mirror twin boundary yielding both 0°+δ and 60°+δ heterostructures separated by the mirror twin boundary (similar to samples in Fig. 2). The 0°+δ heterostructure has triangular domains of alternating conductivity, which we assign to AB and BA crystal structures. The 60°+δ heterostructure has hexagonal domains which we assign to ABBA crystal structure. (c) Closeup CAFM of a single hexagon from the 60°+δ heterostructure showing that the domain walls have regions with low conductivity (gray boxes) and high conductivity (red boxes). Also, the hexagon vertices alternate between low conductivity (gray circles) and high conductivity (red circles). (d) DFT calculations of conduction band minimum (CBM), valence band maximum (VBM), and bandgap for the AB, BA, and ABBA crystal structures. Both the CBM and the VBM of BA lie within the band edges of AB, consistent with the gold domains labeled in (a-b) being more conductive than the green domains for both V$_s$ < 0 and V$_s$ > 0. The order of the bandgaps in increasing order is BA, ABBA, AB. V$_s$ causes fermi level of the AFM tip, E$_{f,tip}$, to translate relative to the heterostructure band edges. (e) Map of wavelength of maximum PL signal for Sample 3 (from Fig. 2) within the range 900 – 1000 nm, which is the expected region for interlayer exciton emission. There is a clear change in emission energy between the 0°+δ heterostructure and the 60°+δ heterostructure. (f) Sample spectrum from the 0°+δ (gold) and 0°+δ (red) heterostructures, showing a clear shift in the PL spectrum. The points for the spectra are labeled in (e).